\documentclass{bmcart}
\usepackage[utf8]{inputenc}
\usepackage{amsthm,amsmath}
\usepackage{amsmath,amssymb,amsfonts}
\usepackage{bm}
\usepackage{cite}
\usepackage{algorithmic}
\usepackage{algorithm,algorithmic}
\usepackage{verbatim}
\usepackage{graphicx} 
\usepackage{epstopdf}
\usepackage{float} 
\usepackage{subfigure} 
\usepackage[utf8]{inputenc} 

\newif\iffiginline
\figinlinetrue

\iffiginline
\else

\def\includegraphics{}
\fi

\startlocaldefs
\endlocaldefs

\begin{document}

\graphicspath{ {./images} }

\begin{frontmatter}

\begin{fmbox}
\dochead{Research}


\title{Optimization of the energy efficiency in Smart Internet of Vehicles assisted by MEC}


\author[
  addressref={aff1},                   
  email={fujfei@seu.edu.cn}   
]{\inits{J.F.}\fnm{Jiafei} \snm{Fu}}
\author[
  addressref={aff1},
  corref={aff1},
  email={\quad \quad p.zhu@seu.edu.cn}
]{\inits{P.Z.}\fnm{Pengcheng} \snm{Zhu}}
\author[
  addressref={aff2},
  corref={aff2},
  email={eehjy@163.com}
]{\inits{J.H.}\fnm{Jingyu} \snm{Hua}}
\author[
  addressref={aff1},
  email={jiaminli@seu.edu.cn}
]{\inits{J.L.}\fnm{Jiamin} \snm{Li}}
\author[
  addressref={aff2},
  email={wjg1214@mail.zjgsu.edu.cn}
]{\inits{J.W.}\fnm{Jiangang} \snm{Wen}}


\address[id=aff1]{
  \orgdiv{National Mobile Communications Research Laboratory},             
  \orgname{Southeast University},          
  \city{Nanjing},                              
  \cny{China}                                    
}
\address[id=aff2]{%
  \orgdiv{School of Information and Electronic Engineering},
  \orgname{Zhejiang Gongshang University},
  \city{Hangzhou},
  \cny{China}
}



\end{fmbox}


\begin{abstractbox}
\begin{abstract} 
Smart Internet of Vehicles (IoV) as a promising application in Internet of Things (IoT) emerges with the development of the fifth generation mobile communication (5G). Nevertheless, the heterogeneous requirements of sufficient battery capacity, powerful computing ability and energy efficiency for electric vehicles face great challenges due to the explosive data growth in 5G and the sixth generation of mobile communication (6G) networks. In order to alleviate the deficiencies mentioned above, this paper proposes a mobile edge computing (MEC) enabled IoV system, in which electric vehicle nodes (eVNs) upload and download data through an anchor node (AN) which is integrated with a MEC server. Meanwhile, the anchor node transmitters radio signal to electric vehicles with simultaneous wireless information and power transfer (SWIPT) technology so as to compensate the battery limitation of  eletric vehicles. 
Moreover, the spectrum efficiency is further improved by multi-input and multi-output (MIMO) and full-duplex (FD) technologies which is equipped at the anchor node. 
In consideration of the issues above, we maximize the average energy efficiency of electric vehicles by jointly optimize the CPU frequency, vehicle transmitting power, computing tasks and uplink rate. Since the problem is nonconvex, we propose a novel alternate interior-point iterative scheme (AIIS) under the constraints of computing tasks, energy consumption and time latency. Results and discussion section verifies the effectiveness of the proposed AIIS scheme comparing with the benchmark schemes. 

\end{abstract}




\end{abstractbox}
%

\end{frontmatter}




\section*{1 Introduction}
\subsection*{\textbf{1.1 Background and related work}}
Recently, with the evolution from the fifth and beyond fifth generation mobile communication (5G/beyond 5G) \cite{b1,b0004} to the sixth generation mobile communication (6G)\cite{b0001}, some novel technologies, such as mobile edge computing (MEC) \cite{b2,b3}, satellite communication \cite{b4} and spatial multiplexing \cite{b5}, have been proposed to promote the development of wireless sensor network (WSN) \cite{b6,b0002}, IoT \cite{b7} and other novel applications \cite{b01,b02,b0003}, so as to realize the interconnection of all things. 
IoV as a special application of IoT attracts extensive attention in terms of road safety, smart transportation and information service \cite{b8}. However, the demand of computing resources and battery capacity becomes more and more urgent due to a large number of data traffic of these novel applications in IoV \cite{b10}. Hence, mobile edge computing is proposed to mitigate the computing burden of electric vehicles \cite{b12}. Meanwhile, the spectrum efficiency is further improved by MIMO and FD technologies \cite{b13}. Besides, the energy consumption of data processing will increase significantly due to the large number of data traffic. Therefore, SWIPT is supposed to be a meaningful method which provides sufficient energy for electric vehicles to complete their intensive computing tasks.

The last decade, mobile cloud computing (MCC) \cite{b14} as a promising method proposed by researcheres to deal with the limited computing resources of vehicles \cite{b15}. However, the network edge meets uncovered service even though massive resources are provided by cloud to afford computing power for vehicles \cite{b16}. In addition, long latency resulting from the long propagation between vehicles and remote access unit are significant \cite{b18}. In consideration of the problems above, MEC technology is proposed to enhance the computing power at the edge of the network \cite{b19,b20,b21}. 
Hence, the author in \cite{b22} investigated the cooperation between MCC and MEC in order to improve the quality of service for vehicles at the network edge. Both \cite{b21} and \cite{b23} aimed at maximizing the system performance and minimizing the time latency assisted by MEC.

Taking account of the energy limitation of electric vehicles, SWIPT \cite{b35} is deployed to extend the battery lifetime of vehicles. Comparing with other practical methods, such as wireless power transfer (WPT) \cite{b002} and energy harvesting (EH) \cite{b32}, SWIPT is used for information decoding and energy harvesting by time-shifting and power splitting, respectively \cite{b2}. Besides, MIMO and FD technologies are able to achieve further improvements of spectral efficiency and low latency \cite{b37,b38}. 

Although considerable research has been devoted to improve performance of vehicular network with MEC-assisted technologies, such as a cooperative computing task and redundant data scheme are  established to reduce the redundancy of data and allocate the computing tasks through MEC server to abtain the minimal system cost \cite{b40}. The author in \cite{b41} developed a software defined network enabled heterogeneous vehicular network to improve the scalability of the network as well as provide high reliability and low-latency communication. K. Zhang, et al, \cite{b42} proposed a task offloading scheme by jointly consider the data transmission and server selection with Q-learning technology. Based on a multi-objective algorithm, an adaptive strategy of offloading is proposed to optimize the resource allocation in \cite{b43}. The author in \cite{b0005} investigated the secrecy energy efficiency in a distributed massive MIMO Systems. In \cite{b001},  an alternative optimization problem is also raised to maximize the transmission rate in a 5G-based IoT system by jointly optimizing the node powers and allocation coefficient.
Rather less attention has been paid to improve energy efficiency of electric vehicles in IoV considering computing tasks, time latency and spectrum efficiency.
\subsection*{\textbf{1.2 Contributions}}
In this paper, we propose an IoV system assisted by MEC which is used for cross-layer offloading to provide low latency and abundant computation resources. Meanwhile, electric vehicles are able to replenish battery capacity by SWIPT technology and the spectrum efficiency is further improved by MIMO and FD technologies. On the premise of computation requirement, time latency and energy constraint of electric vehicles, we maximize the average energy efficiency of electric vehicles. Since the problem is nonconvex, we decouple it into two subproblems at first. According to the previous research \cite{b44}, a closed-form solution of the central processing unit (CPU) frequency is obtained by solving the first sub-problem. Likewise, in order to tackle with the second sub-problem, three sub-problems are divided from the second sub-problem. Finally, an alternate interior-point iterative scheme is proposed to address these subproblems. Numerical simulations are executed to demonstrate the superior performance of our scheme comparing with the benchmark schemes. Our main contributions in this paper can be summarized as follows:
\begin{itemize}
\item We come up an IoV system by using the MEC, the FD and the SWIPT technologies comprehensively, so as to achieve low-latency, abundant computation resources and lower energy consumption by jointly optimizing the CPU frequency, power transfer of electric vehicles, the uplink rate and the offloading tasks.
\item Under the same computation task, energy consumption and time latency constraints, the proposed scheme can yield high energy efficiency, lower time transmission and is able to tackle more computing tasks comparing with the benchmark schemes. 
\item MIMO and FD technologies deployed at anchor node are used to futher improve the spectrum efficiency and the time latency. Meanwhile, we analysis the energy efficiency fairness of the electric vehicles as well.
\end{itemize}
The remainder of this paper is organized as follows. In section 2, we illustrate the system model. A mathematical problem is formulated and solved in section 3. Section 4 provides numerical results and discussions. Finally, section 5 concludes this paper.

\section*{2 Methods}
\subsection*{2.1 System model}
As shown in Fig. \ref{fig1},  we consider an IoV system assisted by MEC, in which a MEC server is deployed at anchor node (AN) and the electric vehicle as a cognitive node connected with AN. Hence, computing tasks will be offloaded partially from electric vehicle nodes (eVNs) to MEC server. Moreover, we assume that the AN works in full-duplex mode which can transmit and receive signals using the same time-frequency resources so as to utilize SWIPT technology to fullfill energy of eVNs as well as receive the offloading computing tasks from the other eVNs at the same time. What's more, each eVN is equipped with a power splitting receiver for ID and EH with the energy conversion ratio $\beta$ ($\beta \in (0,1)$).

In Fig. \ref{fig2}, each vehicle node $VN_i(i \in \{1,...,K\})$ offloads partial computing tasks $m_i(0 \leq m_i \leq M_i)$ to the AN within the time duration $\tau_i^{\text u}$, where $M_i$ denotes the total computation tasks of $VN_i$ and $K$ is the number of vehicle nodes. Simultaneously, AN transmits the computation tasks $m_j^{\text r}(j \in \{1,...,K\})$ executed by MEC server to $VN_j$ within the time duration $\tau_j^{\text d}$, where $m_j^{\text r} = \alpha m_j (0 \leq \alpha \leq 1)$, and $\alpha$ is the ratio of the completed computation tasks from the MEC server. Moreover, the computing time consumption of the MEC server is ignored resulting from the powerful computing ability of it comparing with the vehicle nodes \cite{b45}. Furthermore, the local computing tasks $m_i^{\text {lo}}$, which equals $M_i-m_i$, is completed during the time duration $\tau_i^{\text {lo}}$. Besides, the uplink time consumption of each VN is assumed to be more than the local computing time consumption of that, which is expressed as $0 \leq \tau_i^{\text {lo}} \leq \tau_i$. More details are given as follows.

Notations: The channel fading is modeled as Rayleigh distribution in this paper. $\bm{h}_i^{\text u} \in \mathbb{C c}^{N\times1}$ and $\bm{h}_j^{\text d} \in \mathbb{C c}^{N\times1}$ satisfy uniform distribution$\sim U(0.5,1)$ which present the uplink and downlink channel, respectively, where $N$ is the antenna numbers of AN, and the superscript $\text u$ and $\text d$ refer to uplink and downlink transmission, respectively. Meanwhile, $\bm{H}_{\text {an}}$ represents the self-interference channel of AN. $d_i $ and $\bm{d}_j \in \mathbb{C c}^{N\times1}$ represent the transmitted signal from vehicle node and AN respectively, where $|\bm{d}_j|=1$. Moreover, $p_i$ and $p_j$ denote the transmitting power of VN and AN, respectively. Besides, the additive white Gaussian noise (AWGN) of AN is denoted as $\bm{n}_{\text an}$ with covariance matrix $\sigma_{\text an}^2 \bm{I}_M$. Similarly, $n_j$ and $n_{\text ps}$ are the AWGN of $VN_j$ and PS receiver with variance $\sigma_j^2$ and $\sigma_{\text ps}^2$, respectively. Finally, the bandwidth of the system is expressed as $B$.

\subsubsection*{2.1.1 Computation task processing}
In uplink transmission, partial computing tasks are uploaded from vehicle nodes to the AN in turns. Specifically, $VN_i$ transmits $d_i$ with transmitting power $p_i$ to the AN within the time slot $\tau_i^{\text u}$. Simultaneously, the AN transmits $\bm{d}_j$ with transmitting power $p_j$ to VN. Therefore, we can obtain the received signal at AN, 
\begin{equation}
\begin{aligned}
\bm{x}_i^{u} = \sqrt {{p_i}} {\bm{h}}_i^{\text u}{d_i} + {{\bm{H}}_{\text an}}(\sqrt {{ p_j}} {\bm{d}_j}) + \bm{n}_{\text {an}}
\end{aligned}\label{eq_xi}
\end{equation}
where the first iterm represents the desired signal from $VN_i$, while the self-interference at the AN is expressed as the second iterm, and the rest one is the AWGN of AN. Meanwhile, $p_i$ is the transmitting power of vehicle node which satisfies $p_{\text {min}} \leq p_i \leq p_{\text {max}}$, where $p_{\text {min}}$ and $p_{\text {max}}$ are the minimal and maximal transmitting power of each vehicle node.

According to (\ref{eq_xi}), the received signal of VN is expressed as equation (\ref{eq_Pi}), while equation (\ref{eq_SINRi}) represents the signal to interference and noise ratio (SINR)
\begin{equation}
\begin{aligned}
P_i&=\text{Tr}\{ \bm{x}_i(\bm{x}_i)^H \}\\
&=p_i \text{Tr}\{ \bm{h}_i^{\text u}(\bm{h}_i^{\text u})^H\}+p_j \text{Tr}\{ \bm{H}_{\text {an}}(\bm{H}_{\text {an}})^H\}+\delta_{\text {an}}^2
\end{aligned}\label{eq_Pi}
\end{equation}
\begin{equation}
\begin{aligned}
\gamma_i =\frac{p_i \text{Tr}\{ \bm{h}_i^{\text u}(\bm{h}_i^{\text u})^H\}}{p_j \text{Tr}\{ \bm{H}_{\text {an}}(\bm{H}_{\text {an}})^H\}+\delta_{\text {an}}^2}
\end{aligned}\label{eq_SINRi}
\end{equation}
where $(.)^H$, $\text{Tr}\{.\}$ represent the conjugate and trace of matrix, respectively. According to the equation above, we can obtain the uplink rate which is written as
\begin{equation}
\begin{aligned}
r_i ={\text {B$log_2$}}(1+\gamma_i)
\end{aligned}\label{eq_ri}
\end{equation}

As we know, if the self-interference can be  eliminated completely by self-interference cancelation (SIC), the maximal SINR can be expressed as
 \begin{equation}
\begin{aligned}
\gamma_i^{\text{max}} =\frac{p_i \text{Tr}\{ \bm{h}_i^{\text u}(\bm{h}_i^{\text u})^H\}}{\delta_{\text {an}}^2}
\end{aligned}\label{eq_SIRimax}
\end{equation}
Hence, we can obtain the following ideal uplink rate
\begin{equation}
\begin{aligned}
r_i^{\text{max}} ={\text B} log_2 (1+\gamma_i^{\text{max}})
\end{aligned}\label{eq_rimax}
\end{equation}

However, limited by SIC device, self interference cannot be completely eliminated. Therefore, the uplink rate should not exceed the ideal maximal uplink rate, i.e.,
\begin{equation}
\begin{aligned}
r_i \leq r_i^{\text{max}}
\end{aligned}\label{eq_ricom}
\end{equation}

Based on the analysis above, each time slot $\tau_i^{\text u}$ of uplink transmission and the energy consumption of $VN_i$ can be formulated as follow,
\begin{equation}
\begin{aligned}
\tau_i =\frac{m_i}{r_i}
\end{aligned}\label{eq_ti}
\end{equation}
\begin{equation}
\begin{aligned}
e^{\text{vn}}_i=p_i\frac{ m_i}{r_i}
\end{aligned}\label{eq_evni}
\end{equation}

In consideration of the theory in \cite{b45}, the local computing time duration and the consumption of energy in each vehicle node is written as
\begin{equation}
\begin{aligned}
\tau_i^{\text {lo}}=\sum_{i=1}^{C(M_i-m_i)} \frac{1}{f_i^{\text t}}
\end{aligned}\label{eq_tlo}
\end{equation}
\begin{equation}
\begin{aligned}
e_i^{\text {lo}}=\sum_{n=1}^{C(M_i-m_i)} {\kappa (f_i^{\text t})^2}
\end{aligned}\label{eq_eloi}
\end{equation}
where the required CPU cycle is represented as $C$ which for computing one bit of data. And, for ${\text t}$-th CPU cycles, the CPU frequency of $VN_i$ is $f_i^{\text t}$. Here, $f_i^{\text t}$ should no more than the maximal CPU frequency $f_i^{\text {max}}$. And $\kappa$ denotes the efficient capacitance coefficient in view of the vehicles' chip architecture.  

\subsubsection*{2.1.2 Energy harvesting}
Similarly, in the downlink transmission, $VN_j$ receives signal from the AN including the desired RF signal and the AWGN at $VN_j$, which are expressed as
\begin{equation}
\begin{aligned}
x_j=\sqrt{p_j}(\bm{h}_j^{\text d})^T  \bm{d}_j+n_j
\end{aligned}\label{eq_xj}
\end{equation}

In (\ref{eq_xj}), $(.)^T$ is the transposition operation. Here, the co-channel interference between vehicle nodes are not considered, since the VNs upload the computing tasks to the anchor node in turns. Likewise, we obtain the signal power received at $VN_i$ in (\ref{eq_Pj}) and equation (\ref{eq_SNRj}) is the downlink signal to noise ratio (SNR).
\begin{equation}
\begin{aligned}
P_j={ p_j} \text{Tr}\{(\bm{h}_j^{\text d})^T (\bm{h}_j^{\text d})^{TH}\} +\delta_j^2
\end{aligned}\label{eq_Pj}
\end{equation}
\begin{equation}
\begin{aligned}
\gamma_j =\frac{\beta p_j \text{Tr}\{ (\bm{h}_j^{\text d})^T(\bm{h}_j^{\text d})^{TH}\}}{\delta_j^2}
\end{aligned}\label{eq_SNRj}
\end{equation}

Hence, the downlink rate and the time slot can be derived as
\begin{equation}
\begin{aligned}
r_j =Blog(1+\gamma_j)
\end{aligned}\label{eq_rj}
\end{equation}
\begin{equation}
\begin{aligned}
\tau_j =\frac{\alpha m_j}{r_j}
\end{aligned}\label{eq_tj}
\end{equation}

Based on the coefficient $\beta$ of the PS receiver for EH, the energy of EH at $VN_i$ is shown as
\begin{equation}
\begin{aligned}
e_j^{\text{eh}}= \beta {p_j} \text{Tr}\{(\bm{h}_j^{\text d})^T (\bm{h}_j^{\text d})^{TH}\} \tau_j^{\text d}
\end{aligned}\label{eq_Ehi}
\end{equation}

\subsubsection*{2.1.3 Problem formulation}
The computing time of the MEC server is ignored, since the computing ability of it is more powerful than that of vehicle's  \cite{b45}.
And, we assume that ${\text T}$ is the overall duration. Hence, the relationship between the ${\text T}$ and the total transmission time in uplink and downlink of vehicles as shown in formula (\ref{eq_sumt_i}) and (\ref{eq_sumt_j}), respectively.
 \begin{equation}
\begin{aligned}
0 \leq \sum_{i=1}^K \tau_i^{\text u} \leq {\text T}
\end{aligned}\label{eq_sumt_i}
\end{equation}
\begin{equation}
\begin{aligned}
0 \leq \sum_{j=1}^K \tau_j^{\text d} \leq {\text T}
\end{aligned}\label{eq_sumt_j}
\end{equation}

Moreover, since vehicle node is not able to transmit and receive signal simultaneously, there should be no overlap between uplink and downlink transmission time of a same vehicle node, i.e.,
\begin{equation}
\begin{aligned}
\sum_{i=1}^K \tau_i^{\text u}+ \tau_K^{\text d} \leq {\text T}
\end{aligned}\label{eq_t_i_t_j_1}
\end{equation}
\begin{equation}
\begin{aligned}
\sum_{i=1}^{K_{tmp}} \tau_i^{\text u} \leq {\text T}-\sum_{j=K_{tmp}}^{K} \tau_j^{\text d}
\end{aligned}\label{eq_t_i_t_j_2}
\end{equation}
where $K_{tmp} \in \{1,...,K\}$.

Meanwhile, the downlink transmission slot is less than the uplink transmission slot, so as to gurantee the computing tasks offloading from the next vehicle node can be processed at the AN without time latency and queuing. i.e.,
\begin{equation}
\begin{aligned}
\tau_j^{\text d} \leq \tau_i^{\text u}, j = i-1
\end{aligned}\label{eq_tcom}
\end{equation}

Besides, we assume that each vehicle node has enough energy to complete the computing due to the sufficient energy supply from the AN. Hence, the energy harvested from the AN should no more than the total transmitting power from the AN. Hence, the energy consumption of each vehicle should satisfy,
\begin{equation}
\begin{aligned}
e_i^{\text{vn}} + e_i^{\text {lo}} \leq e_i^{\text{eh}}
\end{aligned}\label{eq_ecom}
\end{equation}
\begin{equation}
\begin{aligned}
e_i^{\text{eh}} \leq e_i^{\text{an}}
\end{aligned}\label{eq_ecom}
\end{equation}

where $e_i^{\text{an}}=p_j \tau_j^{\text d}$. According to the derivations above, we formulate the following problem ($P1$) which is aim to maximizing the average enery efficiency of the total vehicle nodes.
\begin{equation}\label{eq:P1}
    \begin{array}{cl}
      (P1) & \mathop {\max}\limits_{{\bm{f}^{\text t}, \bm{p}, \bm{m},\bm{r}}} \frac{1}{K}\{\sum_{i=1}^K \frac{r_i}{e_i^{\text{vn}}+e_i^{\text {lo}}}\}\\
          & s.t.
	\left \{
		\begin{array}{l}
			C1: e_i^{\text{vn}} + e_i^{\text {lo}} \leq e_i^{\text{eh}}\\
			C2: e_i^{\text{eh}} \leq e_i^{\text{an}}\\
			C3: 0 \leq f_i^{\text t} \leq f_i^{\text{max}}\\
			C4: p_{\text {min}} \leq p_i \leq p_{\text {max}}\\
			C5: 0 \leq m_i \leq M_i\\
			C6: 0 \leq r_i \leq r_i^{\text {max}}\\
			C7: 0 \leq \tau_i^{\text {lo}} \leq \tau_i^{\text u}\\
			C8:	\tau_j^{\text d} \leq \tau_i^{\text u}, j = i-1\\
			C9: 0 \leq \sum_{i=1}^K \tau_i^{\text u} \leq {\text T}\\
			C10: 0 \leq \sum_{j=1}^K \tau_j^{\text d} \leq {\text T}\\
			C11: \sum_{i=1}^{K_{tmp}} \tau_i^{\text u} \leq {\text T}-\sum_{j=K_{tmp}}^{K} \tau_j^{\text d}		
        \end{array} \right.
    \end{array}
\end{equation}
where $\bm{f}^{\text t}=[f_1^{\text t},...,f_K^{\text t}]^T$, $\bm{p}=[p_1,...,p_K]^T$, $\bm{m}=[m_1,...,m_K]^T$, $\bm{r}=[r_1,...,r_K]^T$ represent the CPU frequency, the uplink transmitting power, the offloading computation tasks and the uplink rate of all vehicle nodes, respectively. Moreover, the $f_i^{\text{max}}$ is the maximum CPU frequency.

\subsection*{2.2 Problem solution}
Since the cost function and the constraints in problem ($P1$) are nonlinear and nonconvex, two subproblems are derived from the problem ($P1$) first. 
For the first subproblem, a closed-form expression of $f^{\text t}$ with fixed variables $\bm{p}, \bm{m}, \bm{r}$ is obtained. 
Combining the first subproblem with ($P1$), we derive the second subproblem. For the second subproblem, we utilize an alternate interior-point iterative scheme after decompose it into three subproblems. Finally, the suboptimal varibles $<\bm{f}^{\text t}, \bm{p}, \bm{m}, \bm{r}>$ are abtained. 

\subsubsection*{2.2.1 Local computing optimization}
Inspired by \cite{b45}, the optimal CPU frequency should satisfy
\begin{equation}
\begin{aligned}
f_i^1=f_i^2=...=f_i^{C(M_i-m_i)}=\frac{C(M_i-m_i)}{\tau_i^{\text {lo}}}=\overline f_i
\end{aligned}\label{eq_fn}
\end{equation}
where the average CPU frequency is denoted as $\overline f_i$. 
Based on the derivation above, the original problem ($P1$) is converted into the problem ($P2$) as following.
\begin{equation}\label{eq:P2}
    \begin{array}{cl}
      (P2) & \mathop {\min}\limits_{{\bm{\overline f}}} \frac{1}{K}\{\sum_{i=1}^K e_i^{\text {lo}}\}\\
          & s.t.
	\left \{
		\begin{array}{l}
			C1: e_i^{\text{vn}} + e_i^{\text {lo}} \leq e_i^{\text{eh}}\\
			C2: 0 \leq \tau_i^{\text {lo}} \leq \tau_i^{\text u}\\	
        \end{array} \right.
    \end{array}
\end{equation}
To futher simplify the problem ($P2$), we obtain the following problem ($P3$)
\begin{equation}\label{eq:P3}
    \begin{array}{cl}
      (P3) & \mathop {\min}\limits_{{\bm{\overline f}}} \frac{1}{K}\{\sum_{i=1}^K {\kappa C(M_i-m_i)(\overline{f}_i)^2}\}\\
          & s.t.
	\left \{
		\begin{array}{l}
			C1: e_i^{\text{vn}} + {\kappa C(M_i-m_i)(\overline{f}_i)^2} \leq e_i^{\text{eh}}\\
			C2: 0 \leq \frac{C(M_i-m_i)}{\overline{f}_i} \leq \tau_i^{\text u}\\	
        \end{array} \right.
    \end{array}
\end{equation}
where C2 indicates that $\frac{C(M_i-m_i)}{\tau_i^{\text u}} $ is the lower limit of $\overline{f}_i$ as well as the optimal CPU frequency $f_i^{\text{opt}}$. 

\subsubsection*{2.2.2 Communication optimization}
We obtan the rewritted problem ($P4$) by substituting the optimal CPU frequency $f_i^{\text{opt}}$ into ($P1$). 
\begin{equation}\label{eq:P4}
    \begin{array}{cl}
      (P4) & \mathop {\max}\limits_{{\bm{p}, \bm{m},\bm{r}}} \frac{1}{K}\{\sum_{i=1}^K \frac{r_i}{e_i^{\text{vn}}+e_i^{\text {lo}}}\}\\
          & s.t.
	\left \{
		\begin{array}{l}
			C1: e_i^{\text{vn}} + e_i^{\text {lo}} \leq e_i^{\text{eh}}\\
			C2: e_i^{\text{eh}} \leq e_i^{\text{an}}\\
			C3: p_{\text {min}} \leq p_i \leq p_{\text {max}}\\
			C4: 0 \leq m_i \leq M_i\\
			C5: 0 \leq r_i \leq r_i^{\text{max}}\\
			C6:	\tau_j^{\text d} \leq \tau_i^{\text u}, j = i-1\\
			C7: 0 \leq \sum_{i=1}^K \tau_i^{\text u} \leq {\text T}\\
			C8: 0 \leq \sum_{j=1}^K \tau_j^{\text d} \leq {\text T}\\
			C9: \sum_{i=1}^{K_{tmp}} \tau_i^{\text u} \leq {\text T}-\sum_{j=K_{tmp}}^{K} \tau_j^{\text d}		
        \end{array} \right.
    \end{array}
\end{equation}

Since the problem ($P4$) is still a NP-hard problem, we continue to decompose it into three subproblems, which are denoted as ($P4.1$) $\sim$ ($P4.3$), respectively. 
\begin{equation}\label{eq:P4.1}
    \begin{array}{cl}
      (P4.1) & \mathop {\max}\limits_{{\bm{p}}} \frac{1}{K}\{\sum_{i=1}^K \frac{r_i}{e_i^{\text{vn}}+e_i^{\text {lo}}}\}\\
          & s.t.
	\left \{
		\begin{array}{l}
			C1: e_i^{\text{vn}} + e_i^{\text {lo}} \leq e_i^{\text{eh}}\\
			C2: e_i^{\text{eh}} \leq e_i^{\text{an}}\\
			C3: p_{\text {min}} \leq p_i \leq p_{\text {max}}\\
			C4:	\tau_j^{\text d} \leq \tau_i^{\text u}, j = i-1\\
			C5: 0 \leq \sum_{i=1}^K \tau_i^{\text u} \leq {\text T}\\
			C6: 0 \leq \sum_{j=1}^K \tau_j^{\text d} \leq {\text T}\\
			C7: \sum_{i=1}^{K_{tmp}} \tau_i^{\text u} \leq {\text T}-\sum_{j=K_{tmp}}^{K} \tau_j^{\text d}		
        \end{array} \right.
    \end{array}
\end{equation}

The problem ($P4.1$) is only related to $\bm{p}$, which can be tackled by classical interior-point algorithm.

\begin{equation}\label{eq:P4.2}
    \begin{array}{cl}
      (P4.2) & \mathop {\max}\limits_{{\bm{m}}} \frac{1}{K}\{\sum_{i=1}^K \frac{r_i}{e_i^{\text{vn}}+e_i^{\text {lo}}}\}\\
          & s.t.
	\left \{
		\begin{array}{l}
			C1: e_i^{\text{vn}} + e_i^{\text {lo}} \leq e_i^{\text{eh}}\\
			C2: e_i^{\text{eh}} \leq e_i^{\text{an}}\\
			C3: 0 \leq m_i \leq M_i\\
			C4:	\tau_j^{\text d} \leq \tau_i^{\text u}, j = i-1\\
			C5: 0 \leq \sum_{i=1}^K \tau_i^{\text u} \leq {\text T}\\
			C6: 0 \leq \sum_{j=1}^K \tau_j^{\text d} \leq {\text T}\\
			C7: \sum_{i=1}^{K_{tmp}} \tau_i^{\text u} \leq {\text T}-\sum_{j=K_{tmp}}^{K} \tau_j^{\text d}		
        \end{array} \right.
    \end{array}
\end{equation}

As same as the problem ($P4.1$), interior-point algorithm is used to address the problem ($P4.2$) to achieve the optimal $\bm{m}$.

\begin{equation}\label{eq:P4.3}
    \begin{array}{cl}
      (P4.3) & \mathop {\max}\limits_{{\bm{r}}} \frac{1}{K}\{\sum_{i=1}^K \frac{r_i}{e_i^{\text{vn}}+e_i^{\text {lo}}}\}\\
          & s.t.
	\left \{
		\begin{array}{l}
			C1: e_i^{\text{vn}} + e_i^{\text {lo}} \leq e_i^{\text{eh}}\\
			C2: e_i^{\text{eh}} \leq e_i^{\text{an}}\\
			C3: 0 \leq r_i \leq r_i^{\text{max}}\\
			C4:	\tau_j^{\text d} \leq \tau_i^{\text u}, j = i-1\\
			C5: 0 \leq \sum_{i=1}^K \tau_i^{\text u} \leq {\text T}\\
			C6: 0 \leq \sum_{j=1}^K \tau_j^{\text d} \leq {\text T}\\
			C7: \sum_{i=1}^{K_{tmp}} \tau_i^{\text u} \leq {\text T}-\sum_{j=K_{tmp}}^{K} \tau_j^{\text d}		
        \end{array} \right.
    \end{array}
\end{equation}

Similarly, the optimal variable $\bm{r}$ is obtained by solving the problem ($P4.3$) using the interior-point algorithm.

Each subproblem of ($P4.1$) $\sim$ ($P4.3$) is regarded as an approximate convex problem of the corresponding variable, while the other variable are fixed. In order to solve the problem ($P4$), each subproblem is addressed by classical interior-point algorithm. After that, the whole problem is solved by alternate iterative sheme, which is denoted as \textbf{Algorithm1}. 

\begin{algorithm}
	\renewcommand{\algorithmicrequire}{\textbf{Input:}}
	\renewcommand{\algorithmicensure}{\textbf{Output:}}
	\caption{Alternate interior-point iterative scheme (AIIS)}
	\label{}
	\begin{algorithmic}[1]
		\REQUIRE$\bm{h}_i^{\text u}, \bm{h}_j^{\text d}, \bm{H}_{\text {an}}, \bm{p}_{(0)}, \bm{m}_{(0)}, \bm{r}_{(0)}, \bm{d}_j, d_i, K, N, B, {\text T}, C, \kappa, \alpha, \beta, \delta_j^2, \delta_{\text ps}^2, \delta_{\text {an}}^2, p_{\text {min}}, p_{\text {max}}, M_i, p_{j}$,
 where $\{i,j\}\in \{1,...,K\}$, $\bm{p}_{(0)}$, $\bm{m}_{(0)}$ and $\bm{r}_{(0)}$ represent the initial offloading computation tasks, the uplink transmission power and rate of the vehicle nodes, respectively.
		\STATE According to ($P3$), $f_i^{\text{opt}} = \frac{C(M_i-m_i)}{\tau_i^{\text u}}$, $\tau_i^{\text {lo}}=\tau_i^{\text u}$, and $e_i^{\text {lo}} =\frac{ \kappa C^3 (M_i-m_i)^3}{(\tau_i^{\text u})^2}$.
		\STATE Turn ($P3$) into ($P4$) through substituting $f_i^{opt}=\frac{C(M_i-m_i)}{\tau_i^{\text u}}$, $\tau_i^{\text {lo}}=\tau_i^{\text u}$ and $e_i^{\text {lo}} =\frac{ \kappa C^3 (M_i-m_i)^3}{(\tau_i^{\text u})^2}$ into ($P3$).
		\STATE  Let $\bm{p}_{ini}=\bm{p}_{(0)}, \bm{m}_{ini}=\bm{m}_{(0)}, \bm{r}_{ini}=\bm{r}_{(0)}$, and iteration number $k=1$ be the initial value.
		\REPEAT
		\STATE Deal with ($P4.1$) with fixed $\bm{m}_{(k-1)}, \bm{r}_{(k-1)}$, $\bm{p}_{ini}=\bm{p}_{(k-1)}$, to abtain $\bm{p}_{(k)}^{(\text{opt})}$;
		\STATE Deal with ($P4.2$) with fixed $\bm{p}_{(k-1)}, \bm{r}_{(k-1)}$, $\bm{m}_{ini}=\bm{m}_{(k-1)}$, to abtain $\bm{m}_{(k)}^{(\text{opt})}$;
		\STATE Deal with ($P4.3$) with fixed $\bm{p}_{(k-1)}, \bm{m}_{(k-1)}$, $\bm{r}_{ini}=\bm{r}_{(k-1)}$, to abtain $\bm{r}_{(k)}^{(\text{opt})}$;
		\STATE Let $k=k+1$;
		\UNTIL the objective function in ($P4$) converges.
		\ENSURE $  \bm{f}^{(\text{opt})}, \bm{p}^{(\text{opt})}, \bm{m}^{(\text{opt})}, \bm{r}^{(\text{opt})}$ 
	\end{algorithmic}
\end{algorithm}

\section*{4 Results and discussion}
Based on numerical simulations, we dicuss the performance of the proposed scheme in this papaer comparing with the two benchmark schems. Three offloading strategies are denoted as below and the simulation parameters are listed in \textbf{Table 1}.
\begin{itemize}
\item FVS: all variables are fixed.
\item FOS: computing tasks completely offload to the AN which indicates the local computing task is zero.
\item AIIS: computing tasks are offloaded arbitrarily based on the maximum average energy efficiency of total vehicle nodes. 
\end{itemize}

In Fig. \ref{fig3}, we find that the average energy efficiency of the total vehicles decrease as computing tasks increase. However, the proposed AIIS scheme still has the maximum average energy efficiency comparing with the benchmark schemes. 
Fig. \ref{fig4} investigates the influence of antenna numbers on system performance. Here, we assume that the number of vehicles equals 10 and the antenna numbers are \{6,8,10\}. As shown in figure, with the number of antennas increasing, the average energy efficiency for each schemes is improved. However, the average energy efficiency of the proposed scheme is higher than that of the comparing schemes with the same antenna numbers. 

Fig. \ref{fig5} studies the system performance with different number of vehicle nodes from 1 to 10, where we can find that the number of vehicles make no difference to the average energy efficiency for each schemes due to the total energy efficiency of all vehicles are averaged. Nerverthless, the proposed scheme is superior to the comparing scheme under the same number of vehicles. 

Fig. \ref{fig6} studies the effect of the downlink transmitting power from the AN. Generally speaking, additional downlink transmitting power means that vehicle node may have more energy to complete the computation tasks. However, since each vehicle node tends to maximize its uplink rate as well as minimize its uplink energy consumption so as to maximize the average energy efficiency. Hence, the redundant downlink transmitting power seems to have no effect on the average energy efficiency. 
In other words, we may achieve lower energy consumption with higher energy efficiency. Besides, the system performance of the proposed scheme is better than that of the other two schemes as well.   

Under the same coditions, Fig. \ref{fig7} shows that the transmission time of three schemes increase with the number of computation tasks increasing. However, the proposed scheme has the lower transmission time compared to the other schemes.

To futher study the fairness of each vehicles, we obtain the Fig. \ref{fig8} and Fig. \ref{fig9}. For the former, variance of energy efficiency of all vehicles are given to prove that the proposed scheme AIIS has the best energy efficiency fairness between the same vehicle numbers compared to the other schemes, which indicates that the qulity of service of different vehicles can be guranteed. 
For the latter, the proposed scheme has the lower variance of transmission time consumptions between all vehicles with the same computation tasks comparing with the benchmark schemes, which indicates that the time-latency fairness between the different vehicles of AIIS is superior to that of the comparing schemes. 

\section*{5 Conclusion}
In this paper, we propose an IoV system assisted by MEC server which is deployed at anchor node. Electric vehicle as a cognitive node uploads their intensive computing tasks to the AN as well as harvests energy from the RF signal transmitted by the AN with SWIPT technology, so as to alleviate the heavy computing tasks, reduce the time latency and compensate the limited battery capacity of vehicle node. Besides, the spectral efficiency is further improved by MIMO and FD technologies. Finally, an alternate interior-point iterative scheme (AIIS) is proposed to deal with a non-convex problem which is aim to maximize the average energy efficiency of vehicles by jointly optimize the computing and communication resources. Simulation results verify that the proposed AIIS scheme outperforms the other two comparison schemes. Furthermore, the service of future IoV system may benefit from the proposed scheme.





\begin{backmatter}

\section*{Acknowledgements}
Not applicable.

\section*{Funding}
This work was supported by the National Key R\&D Program of China under Grant 2021YFB2900300, and the National Natural Science Foundation of China under Grants 62171126 and 61871122.\\
Zhejiang Natural Science Foundation (LQ21F010008).

\section*{Abbreviations}
IoV: Internet of Vehicle; IoT: Internet of Things; 5G: the fifth generation mobile communication; 6G: the sixth generation of
mobile communication; MEC: mobile edge computing; eVNs: electric vehicle nodes; AN: anchor node; SWIPT: simultaneous
wireless information and power transfer; FD:full-duplex; MIMO: multi-input and multi-output; WSN: wireless sensor
network; MCC: mobile cloud computing; EH: energy harvesting; WPT: wireless power transfer; ID: information decoding; PS: power-splitting; TS: timeswitching; SDN: software defined network; CPU: central processing unit; SINR: interference plus noise ratio; SIC: selfinterference cancelation; RF: radio frequency; SNR: signal to noise.



\section*{Competing interests}
The authors declare that they have no competing interests.


\section*{Authors' contributions}
Jiafei Fu proposes an alternate interior-point iterative scheme (AIIS) to optimize the energy efficiency of electric vehicles in a smart IoV system and she is the major contributor in writing the manuscript. All authors read and approved the final manuscript.

\section*{Authors' information}
\textbf{Jiafei Fu} received her B.S. degree in electronic and information engineering from Lishui University, Lishui, China, in 2017 and the M.S. degree in information and communication engineering from Zhejiang University of Technology, Hangzhou, China, in 2020. She is currently pursuing the Ph.D degree in information and communication engineering at Southeast University, Nanjing, China. Her research interests include full-duplex, distributed MIMO, mobile edge computing, and ultra reliable and low latency communication.\\

\textbf{Pengcheng Zhu} (M'09) received his B.S. and M.S. degrees in electrical engineering from Shandong University, Ji'nan, China, in 2001 and 2004, respectively, and his Ph.D. degree in information and communication engineering from Southeast University, Nanjing, China, in 2009. He is a professor with the National Mobile Communications Research Laboratory, Southeast University. His research interests lie in the areas of wireless communications and mobile networks, including B5G/6G mobile communication systems, physical layer security and privacy, distributed MIMO, and mmWave communications.\\

\textbf{Jingyu Hua} was born in Zhejiang province, China in 1978. He received the B.S. and M.S. degrees in electronic engineering from the South China University
of Technology, Guangzhou, China, in 1999 and 2002. Then in 2006, he received the Ph.D. degree in electronic engineering from Southeast University, Nanjing, China. Since 2006, he had joined Zhejiang University of Technology as an assistant professor in the electronic engineering department, and promoted as full professor in 2012. From 2019, he is with Zhejiang Gongshang University as a distinguish professor. He is the author of more than 200 articles and more than 20 inventions. His research interests include the area of parameter estimation, channel modeling, wireless localization and digital filtering in wireless communications. He is currently an associate editor for the IEEE Transactions on Instrumentation and Measurements.\\

\textbf{Jiamin Li} received the Ph. D. degree in electronic engeneering from Southeast University, Nanjing, China in 2014. He joined the National Mobile Communications Research Laboratory at Southeast University, China, in 2015, where he has been an Associate Professor since 2018. His current research interests include turbo detection, vehicular communications and massive MIMO systems.

\textbf{Jiangang Wen} was born in 1989. He received the B.S. degree of electronic and information engineering in 2012, and the Ph.D. degree of Control Science and Engineering in 2019, both from the Zhejiang University of Technology, Hangzhou, China. Currently, he is a teacher at Zhejiang Gongshang University. His research interests include optimization and digital filtering in mobile communications.\\


\bibliographystyle{bmc-mathphys} 
\bibliography{bmc_article}      




\section*{Figures}
\begin{figure}[h!]
\iffiginline
\includegraphics[width=0.8\textwidth]{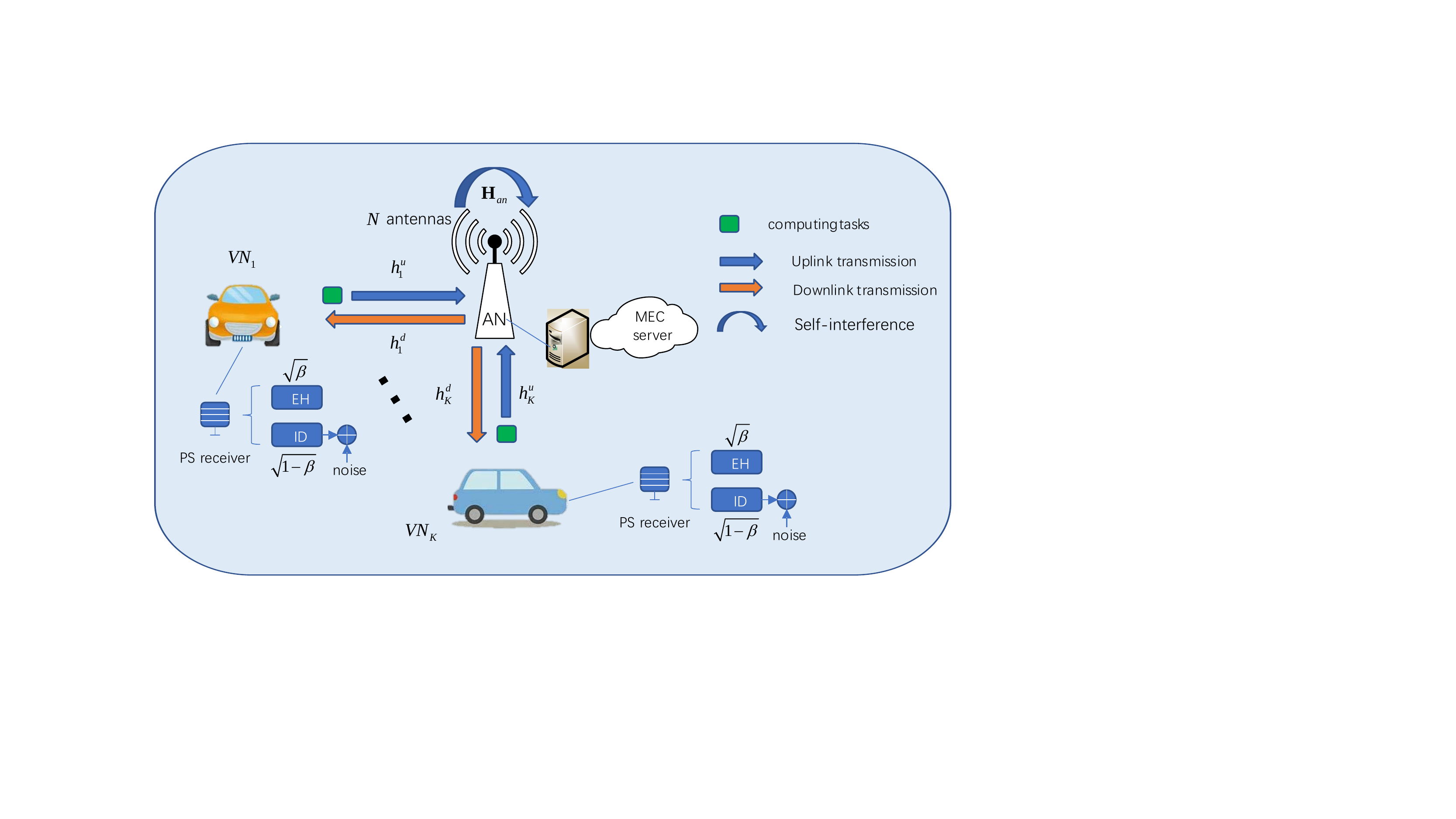}
\fi
 \caption{System model}
\label{fig1}
\end{figure}

\begin{figure}[h!]
\iffiginline
\includegraphics[width=0.8\textwidth]{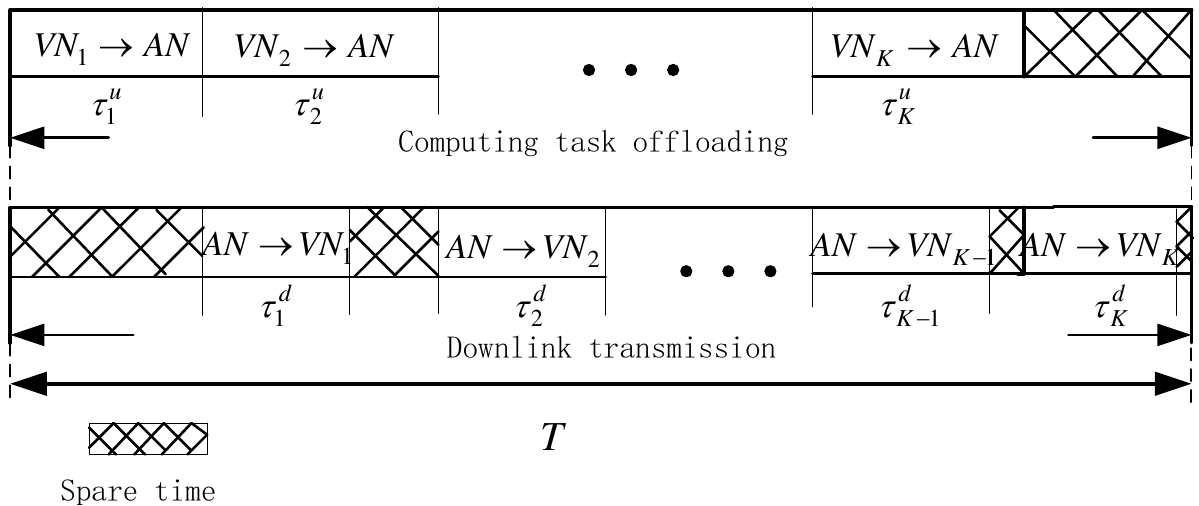}
\fi
 \caption{Time slot}
\label{fig2}
\end{figure}

\begin{figure}[h!]
\iffiginline
\includegraphics[width=0.8\textwidth]{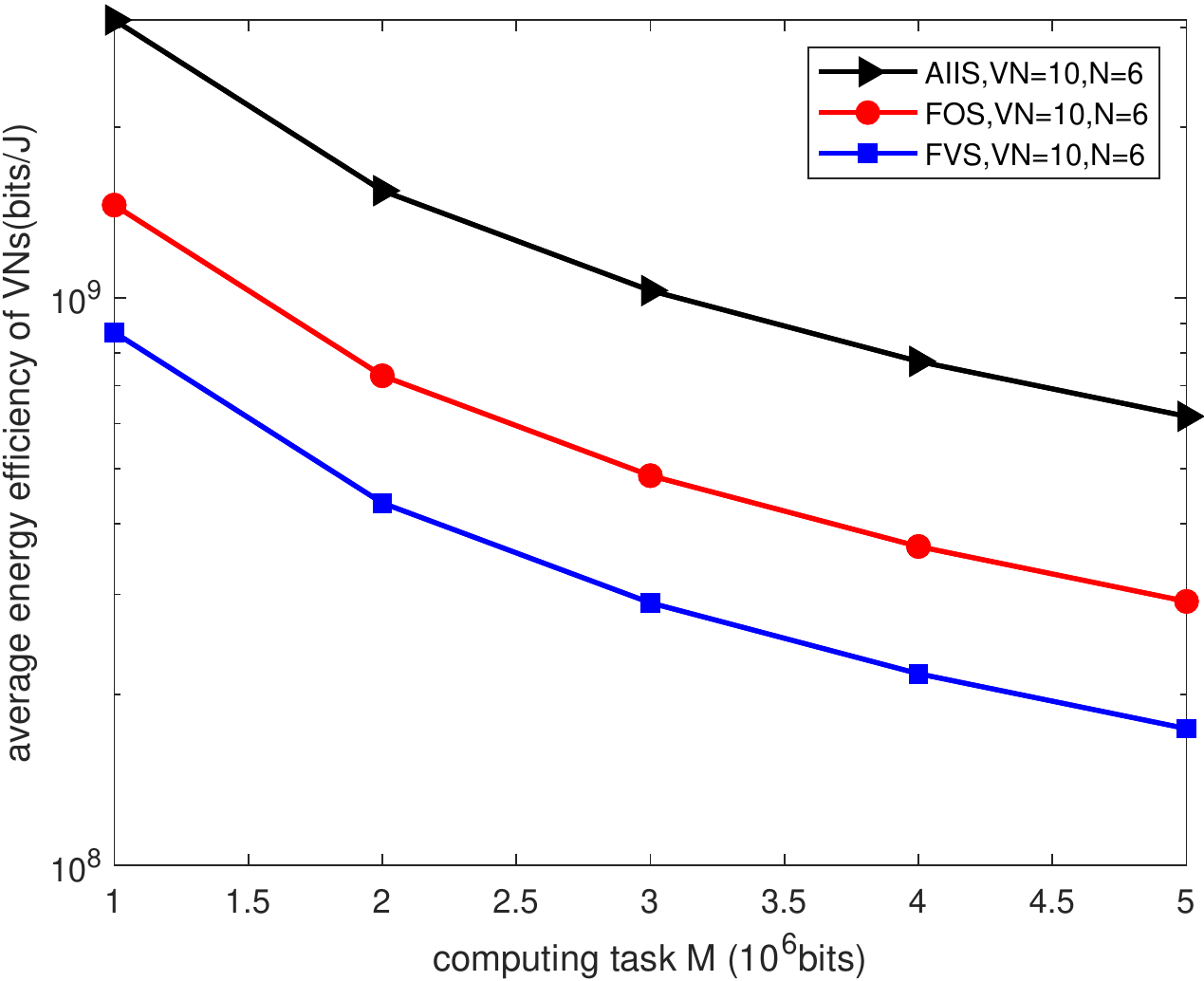}
\fi
 \caption{The average energy efficiency of vehicles versus the
computation task M}
\label{fig3}
\end{figure}

\begin{figure}[h!]
\iffiginline
\includegraphics[width=0.8\textwidth]{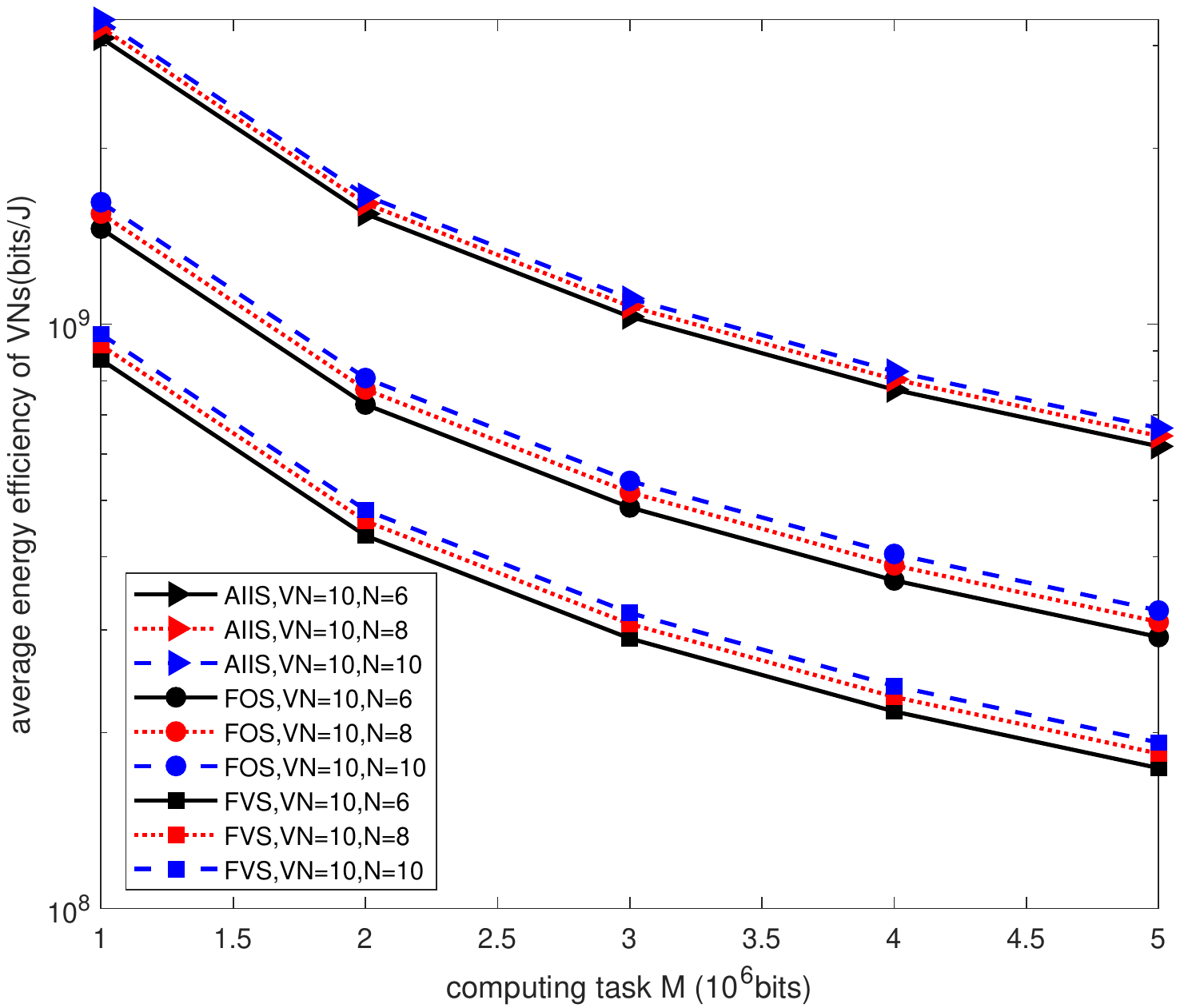}
\fi
 \caption{The average energy efficiency of vehicle nodes (VNs) versus the
computation task M under different antenna numbers N}
\label{fig4}
\end{figure}

\begin{figure}[h!]
\iffiginline
\includegraphics[width=0.8\textwidth]{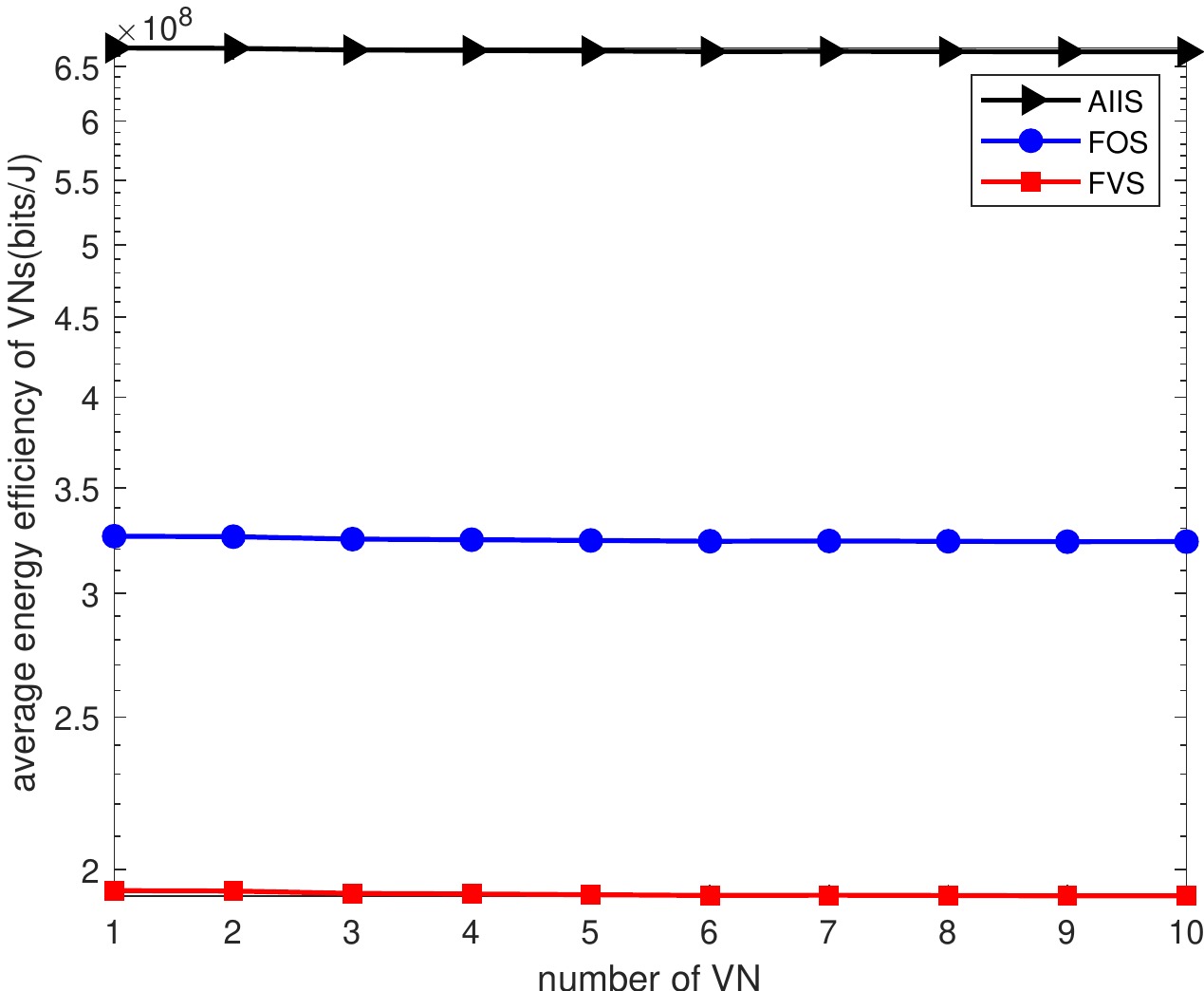}
\fi
 \caption{The average energy efficiency of vehicle nodes (VNs) versus the
different numer of VNs}
\label{fig5}
\end{figure}

\begin{figure}[h!]
\iffiginline
\includegraphics[width=0.8\textwidth]{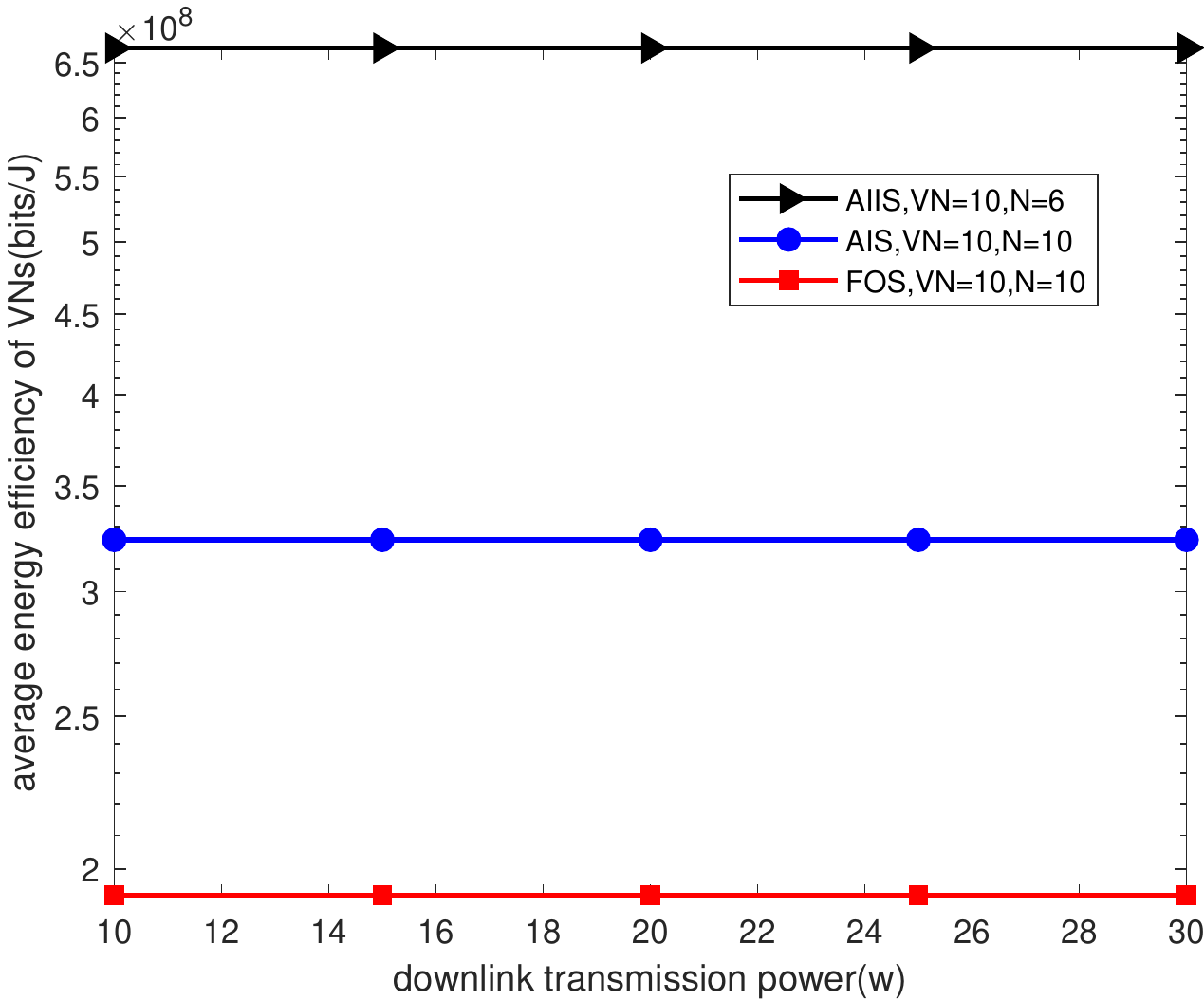}
\fi
 \caption{The average energy efficiency of vehicle nodes (VNs) versus the
different transmitting power of AN}
\label{fig6}
\end{figure}

\begin{figure}[h!]
\iffiginline
\includegraphics[width=0.8\textwidth]{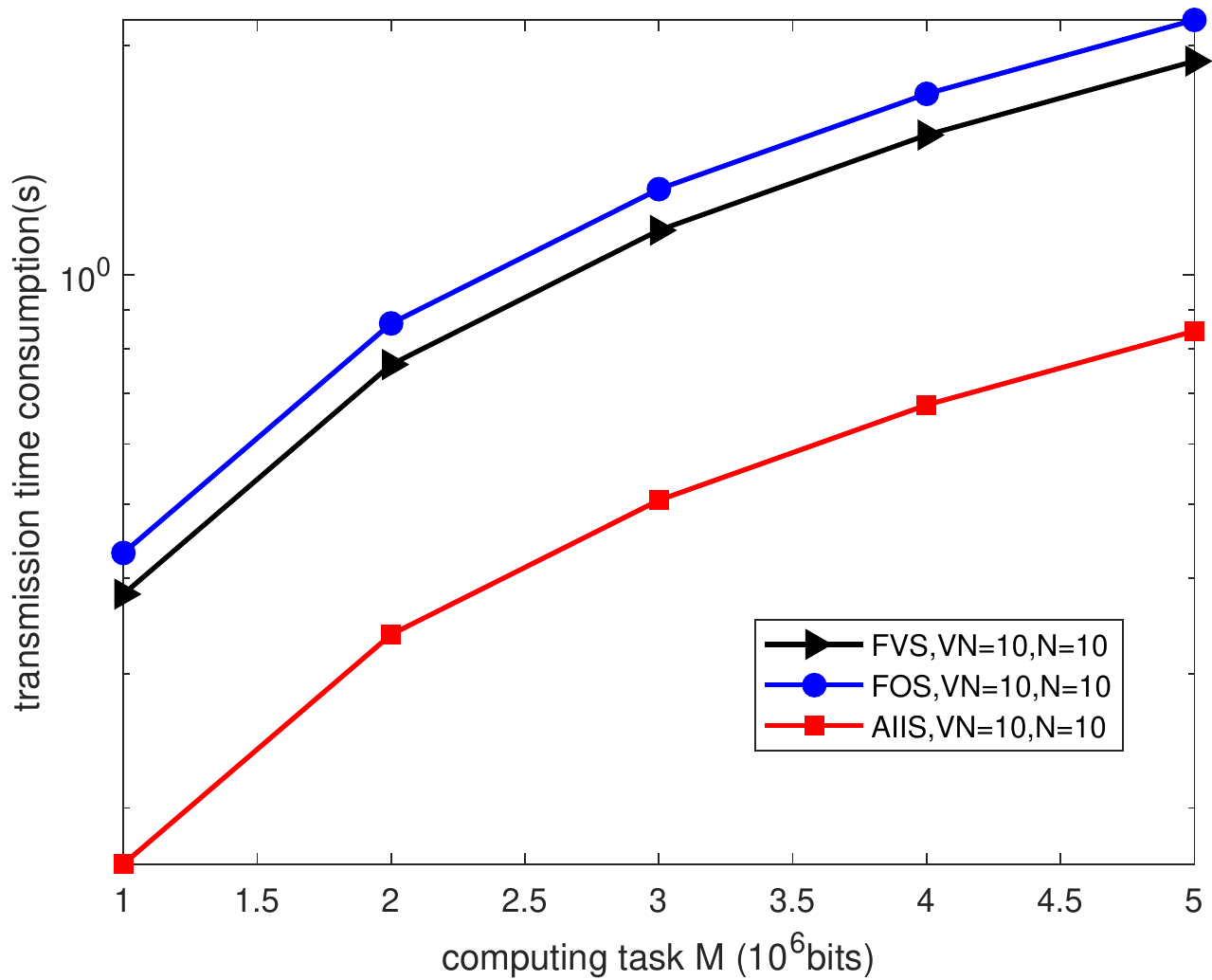}
\fi
 \caption{The transmission time consumption of vehicle nodes (VNs) versus the
computation task M}
\label{fig7}
\end{figure}

\begin{figure}[h!]
\iffiginline
\includegraphics[width=0.8\textwidth]{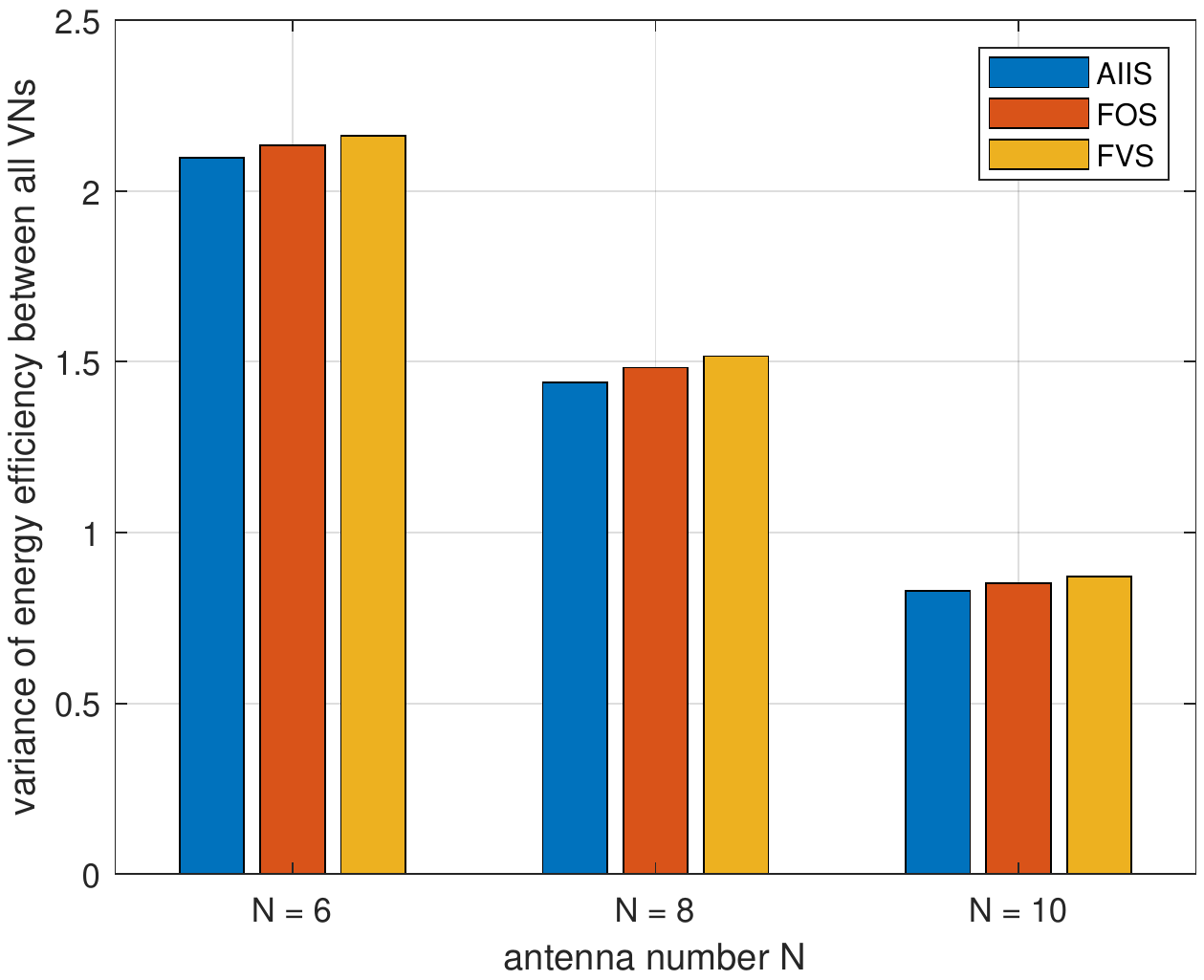}
\fi
 \caption{The variance of energy efficiency between all vehicle nodes (VNs) under different antenna numbers N}
\label{fig8}
\end{figure}

\begin{figure}[h!]
\iffiginline
\includegraphics[width=0.8\textwidth]{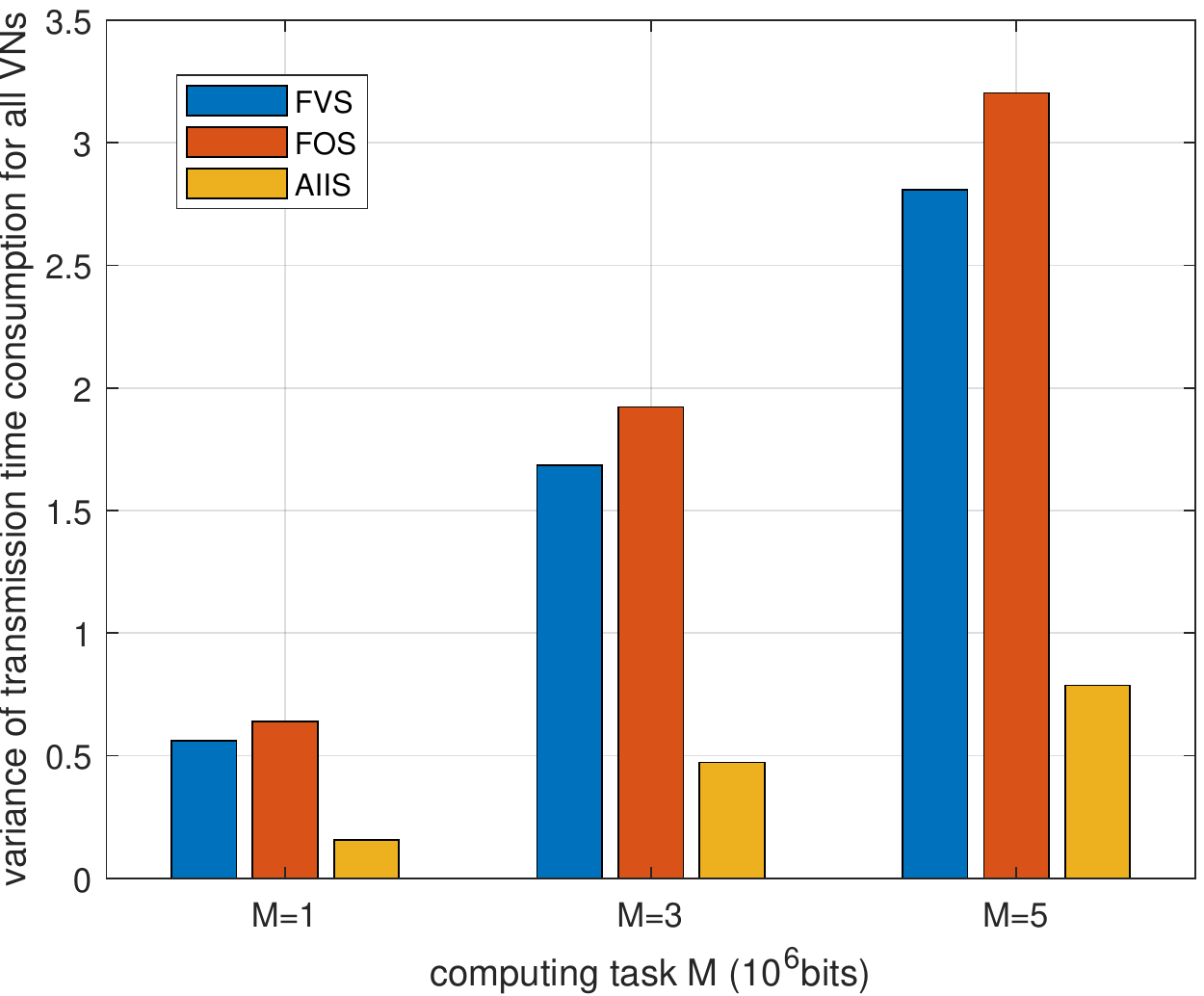}
\fi
 \caption{The variance of transmission time consumption of vehicle nodes (VNs) under different computation task M}
\label{fig9}
\end{figure}


\section*{Tables}
\begin{table}[h!]
\caption{Simulation parameters}
  \begin{tabular}{|c | c|}
    \hline
parameter& value\\
\hline
$\bm{h}_i^{\text u}, \bm{h}_j^{\text d}$&
Random uniform distribution $ \sim U(0.5,1)$\\
$\bm{H}_{\text {an}}$&
Raylaigh Fading distribution $ \sim \mathcal{CN}(0,0.1)$\\
$N$ &
$6,8,10$ \\
$K$&
$1 \sim 10$ \\
$B$&
2MHz \\
$C$&
$10^3cycles/bit$ \\
$T$&
0.5\\
$\beta$&
0.2 \\
$\alpha$&
0.8 \\
$\kappa$&
$10^{-33}$ \\
$p_j$&
$10W\sim30W$ \\
$p_{\text {min}}$&
$1W$ \\
$p_{\text {max}}$&
$5W$ \\
$\delta_{\text an}^2, \delta_j^2, \delta_{\text ps}^2$&
$10^{-7}W$ \\
\hline
  \end{tabular}
\end{table}

\end{backmatter}
\end{document}